\documentstyle[preprint,aps]{revtex}

\begin{document}
\author{Amit Keren and Ophir M. Auslaender}
\title{Spin Echo Decay in a Stochastic Field Environment }
\address{Department of Physics, Technion - Israel Institute of Technology, \\
Haifa 32000, Israel.}
\maketitle

\begin{abstract}
We derive a general formalism with which it is possible to obtain the time $%
(\tau )$ dependence of the echo size for a spin in a stochastic field
environment. Our model is based on ``strong collisions''. We examine in
detail three cases where: (I) the local field is $\pm \omega _{0}$, (II) the
field distribution is continuous and has a finite second moment, and (III)
the distribution is Lorentzian. The first two cases show a $T_{2}$ minimum
effect and are exponential in $\tau ^{3}$ as $\tau \rightarrow 0$. The last
case can be approximated by the phenomenological expression $\exp (-[2\tau
/T_{2}]^{\beta })$ with $1<\beta <2$, where in the $\tau \rightarrow 0$
limit $\beta =2$.
\end{abstract}

\newpage

Spin echo decay (SED) measurements, also known as $T_{2}$, are conducted by
a variety of experimental techniques, such as RF-$\mu $SR \cite
{KreitzmanPRL88}, ESR \cite{ESR}, NQR, and NMR \cite{NMR}. With the recent
explosion of high-$Tc$ superconductivity research, NMR-$T_{2}$ measurements
in particular are receiving renewed attention, since they are very
successful in probing both the normal \cite{PenningtonPRB89} and
superconducting states\ \cite{CarrettaPRB92} of cuprates. These experiments
lead to a revival of theoretical activity, focusing on the calculation of
the SED waveform for different sources of interactions such as spin lattice
coupling, spin-spin coupling, and stochastic fluctuations. For this purpose,
a variety of analytical \cite{AnsermetJCP88} and numerical \cite
{WalstedetPRB95} models were applied. However, several dynamical features,
observed experimentally, have not been accounted for. In this paper we
provide new insight into these features by re-examining the echo decay
waveform of a spin in a stochastic field environment, and use an analytical
approach based on the ``strong collision'' model (see below) to yield
quantitative understanding of SED.

An earlier exact treatment of the stochastic problem, based on a diffusion
like model, was presented by Klauder and Anderson (KA) \cite{KlauderPR62}.
They found that for Lorentzian diffusion the waveform is Gaussian, and for
Gaussian diffusion the waveform is exponential in $\tau ^{3}$ (see bellow).
Although the KA approach is physically more intuitive, the final result
lacks three features: (I) the waveform does not depend on the diffusion
rate, and therefore it cannot change continuously (for example, as a
function of temperature), (II) they could not account for stretched
exponential relaxation $\exp (-[2\tau /T_{2}]^{\beta })$ with $\beta <2$,
and (III) the SED depends monotonically on the diffusion rate, although it
is natural to expect that when the diffusion is either very fast or
extremely slow, the echo does not decay. As we shall see, our derivation
allows for all these phenomena, and therefore might be applicable to some
cases to which the diffusion model is not. In addition, we examine one case
which was not considered by KA, namely, the local field is $\pm \omega _{0}$%
. For this case our derivation is exact.

In echo NMR, NQR, and ESR transverse relaxation measurements, a $\pi /2$
pulse is applied to a system of spins polarized along the $z$ direction. As
a result, a net polarization along the $x$ direction ($M_{x}$) is obtained.
In RF-$\mu $SR the muons enter the sample with their spin already polarized
along the $x$ direction. After the pulse (or muon arrival), the spins evolve
with time, each one in its local field $B_{z}$, until time $\tau $ when a $%
\pi $ pulse is applied, sending the $x$ component of each spin $S_{x}$ to $%
-S_{x}$ (and $S_{z}$ to $-S_{z}$). The spins then continue to evolve, and if 
$B_{z}$ is static, an echo is formed at time $2\tau $. If, however, the
local field is dynamic, the phase acquired by the spin before the $\pi $
pulse is not necessarily equal to the phase lost after it, and the echo size
diminishes as a function of $\tau $. This situation can be quantified by 
\begin{equation}
M_{x}(2\tau )=M_{x}(0)\left\langle \cos \left[ \int_{0}^{\tau }\omega
(t)dt-\int_{\tau }^{2\tau }\omega (t)dt\right] \right\rangle ,
\label{CompliteM}
\end{equation}
where $\omega (t)=\gamma B_{z}(t)$, $\gamma $ is the spin's gyromagnetic
ratio, and $\left\langle {}\right\rangle $ is an average over all possible
frequency trajectories.

First we would like to evaluate Eq.~\ref{CompliteM} to lowest order in $\tau 
$. Assuming that the argument of the cosine is small, we can expand it to
second order, and then evaluate terms such as $\int_{0}^{\tau
}\int_{0}^{\tau }dt^{\prime }dt^{\prime \prime }\left\langle \omega
(t^{\prime })\omega (t^{\prime \prime })\right\rangle $ and $\int_{0}^{\tau
}\int_{\tau }^{2\tau }dt^{\prime }dt^{\prime \prime }\left\langle \omega
(t^{\prime })\omega (t^{\prime \prime })\right\rangle $. Assuming a
correlation function of the form 
\begin{equation}
\left\langle \omega (t^{\prime })\omega (t^{\prime \prime })\right\rangle
=\left\langle \omega ^{2}\right\rangle \exp \left( -\nu \left| t^{\prime
\prime }-t^{\prime }\right| \right) =\left\langle \omega ^{2}\right\rangle
\left( 1-\nu \left| t^{\prime \prime }-t^{\prime }\right| +\ldots \right) ,
\label{Correlations}
\end{equation}
where $\left\langle \omega ^{2}\right\rangle $ is the second moment of the
instantaneous field distribution, we find 
\begin{equation}
M_{x}(2\tau )=M_{x}(0)\left( 1-\frac{2}{3}\left\langle \omega
^{2}\right\rangle \nu \tau ^{3}+\ldots \right) .  \label{EarlyTime}
\end{equation}
Equation \ref{EarlyTime} is well known \cite{NMR} and will serve as a test
of our derivation.

Next we shall evaluate Eq.~\ref{CompliteM} to all orders in $\tau $ by
making some assumptions concerning $\omega (t)$. We quantify the dynamical
fluctuation using ``indirect echo'' and the strong collision model. Indirect
echo is equivalent to the situation described by Eq.~\ref{CompliteM} but
instead of $S_{x}\rightarrow -S_{x}$ at the $\pi $ pulse, the frequency is
reversed ($\omega \rightarrow -\omega $); in Fig.~\ref{EchoDemo}a we
demonstrate indirect echo by showing that a reversal of $\omega $ at $\tau $
leads to $S_{x}(2\tau )=S_{x}(0)\equiv 1$. The strong collision model
accounts for $\omega (t)$ by allowing frequency changes only at specific
times $t_{1},t_{2}\ldots t_{n}$. The probability density of finding the
frequency $\omega $ at any time interval is taken to be the line shape $\rho
(\omega )$. A demonstration of this situation for a particular spin is
presented in Fig.~\ref{EchoDemo}b. Here the spin has experienced two
frequency changes at times $t_{1}$ and $t_{2}$ before the $\pi $ pulse and
one change after the $\pi $ pulse at $t_{3}$. As a result $S_{x}(2\tau )\neq
S_{x}(0)$ and on average the echo size will decrease as a function of $\tau $%
. This type of dynamical process results in a correlation function in the
form of Eq.~\ref{Correlations}. By comparison, in KA's model the frequency
after each change depends on the frequency before the change.

We shall now treat the case of an ensemble of spins and average over all
possible field changes, the times at which they take place, and all possible
fields in each time interval. If there are $n$ hops at times $t_{1,}\ldots
,t_{n}$ before the $\pi $ pulse and $m$ hops at times $t_{n+1},\ldots
,t_{m+n}$ between the $\pi $ and the observation time $t=2\tau $, the phase
acquired by the spin ($\theta _{n,m}$) is 
\begin{eqnarray}
\theta _{n,m} &=&\omega _{n+m+1}(t-t_{n+m})+\sum_{j=2}^{m}\omega
_{j+n}(t_{n+j}-t_{n+j-1}) \\
&-&\omega _{n+1}(t_{n+1}-\tau )+\omega _{n+1}(\tau
-t_{n})+\sum_{i=1}^{n}\omega _{i}(t_{i}-t_{i-1}).  \nonumber
\end{eqnarray}
The polarization along the $x$ axis is therefore 
\[
M_{x}(\omega _{1},\ldots ,\omega _{n+m+1};t,\tau ;t_{1},\ldots
,t_{n+m})\equiv \Re \exp (i\theta _{n,m}), 
\]
where $\Re $ stands for the real part; we shall omit it from now on. We
first average over all possible frequencies $\omega _{i}$ in the time
segment $[t_{i-1},t_{i}]$ and define 
\[
M_{x}(t,\tau ;t_{1},\ldots ,t_{n+m})\equiv \int \rho (\omega _{1})d\omega
_{1}\ldots \int \rho (\omega _{n+m+1})d\omega _{n+m+1}M_{x}(\omega
_{1},\ldots ,\omega _{n+m+1};t,\tau ;t_{1},\ldots ,t_{n+m+1}). 
\]
This results in 
\[
M_{x}(t,\tau ;t_{1},\ldots ,t_{n+m})=g(t-t_{n+m})\left[
\prod_{j=2}^{m}g(t_{j+n}-t_{j+n-1})\right] g(2\tau -t_{n+1}-t_{n-1})\left[
\prod_{i=1}^{n}g(t_{i}-t_{i-1})\right] 
\]
where $g(t)$, also known as the free induction decay (FID) function, is
given by 
\[
g(t)=\int_{-\infty }^{\infty }\rho (\omega )\exp (i\omega t)d\omega . 
\]
The probability density of finding exactly $n+m$ hops at times $t_{1},\ldots
,t_{n+m}$ is 
\[
\exp \left[ -\nu (t-t_{n+m})\right] \prod_{i=1}^{n+m}\exp \left[ -\nu
(t_{i}-t_{i-1})\right] \nu dt_{i}=\nu ^{n+m}\exp (-\nu
t)\prod_{i=1}^{n+m}dt_{i}, 
\]
where $\nu $ is the field selection rate. Thus, the averaged spin
polarization at time $t$ is given by 
\begin{equation}
M_{x}(t)=\sum_{n=0}^{\infty }\sum_{m=0}^{\infty }\nu ^{n+m}\exp (-\nu
t)I_{n,m}(t,\tau ),  \label{AnExpresionForM}
\end{equation}
where 
\begin{equation}
I_{n,m}(t,\tau )=\int_{\tau }^{t}dt_{n+m}\cdots \int_{\tau
}^{t_{n+2}}dt_{n+1}\int_{0}^{\tau }dt_{n}\cdots
\int_{0}^{t_{2}}dt_{1}M_{x}(t,\tau ;t_{1},\ldots ,t_{n+m}).
\label{Integral1}
\end{equation}
The integration limits guarantee that $t_{i+1}>t_{i}$.

We can simplify Eq.~\ref{Integral1} by turning the time at which the $\pi $
pulse is applied $(\tau )$ into a running variable $(t^{\prime })$ whose
value is fixed with a $\delta $ function. The $\delta $ function should
force the sum of time segments from zero until $\tau $ to be equal to the
sum of time segments from the $\tau $ until $2\tau $, namely, 
\[
\delta (t^{\prime }-\tau )=2\delta \left( (t^{\prime
}-t_{n})+\sum_{i=1}^{n}(t_{i}-t_{i-1})-%
\sum_{j=2}^{m+1}(t_{n+j}-t_{n+j-1})-(t_{n+1}-t^{\prime })\right) 
\]
where $t_{n+m+1}$ stands for $2\tau $. As a result 
\begin{equation}
I_{n,m}(2\tau ,\tau )=\int_{0}^{2\tau }dt_{n+m}\cdots
\int_{0}^{t_{n+2}}dt_{n+1}\int_{0}^{t_{n+1}}dt^{\prime }\int_{0}^{t^{\prime
}}dt_{n}\cdots \int_{0}^{t_{2}}dt_{1}M_{x}(2\tau ,t^{\prime };t_{1}\ldots
t_{n+m})\delta (t^{\prime }-\tau ),  \label{Integral2}
\end{equation}
and the integrand in Eq.~\ref{Integral2} is a function of time differences
only.

We now introduce the integral representation of the $\delta $ function 
\begin{equation}
\delta (x)=\frac{1}{2\pi }\int_{-\infty }^{\infty }\exp (i\Omega x)d\Omega ,
\label{DeltaDef}
\end{equation}
and the Laplace transform of $M_{x}$: 
\begin{equation}
\overline{M}_{x}(s)=2\int_{0}^{\infty }M_{x}(2\tau )\exp (-2s\tau )d\tau .
\label{LaplaceDef}
\end{equation}
By inserting Eq.~\ref{DeltaDef} into Eq.~\ref{Integral2}, Eq.~\ref{Integral2}
into Eq.~\ref{AnExpresionForM} and substituting this in Eq.~\ref{LaplaceDef}
we find that all the integrals decouple and 
\begin{equation}
\overline{M}_{x}(s)=\frac{1}{2\pi }\int_{-\infty }^{\infty }d\Omega
f_{2}(z_{-},z_{+})\sum_{n=0}^{\infty }\sum_{m=0}^{\infty }\left( \nu
f_{1}(z_{-})\right) ^{n}\left( \nu f_{1}(z_{+})\right) ^{m}  \label{SumDef}
\end{equation}
where 
\begin{equation}
z_{\pm }=s+\nu \pm i\Omega /2,  \label{DefOfZ}
\end{equation}
\begin{equation}
f_{1}(z_{\pm })=\int_{0}^{\infty }du\exp \left( -z_{\pm }u\right) g(u),
\label{f1Def}
\end{equation}
and 
\begin{equation}
f_{2}(z_{-},z_{+})=\frac{f_{1}(z_{-})+f_{1}(z_{+})}{z_{-}+z_{+}}.
\label{f2Def}
\end{equation}
Finally, $\left| \nu f(z)\right| <1$, and performing the sum in Eq.~\ref
{SumDef} gives

\begin{equation}
\overline{M}_{x}(s)=\frac{1}{2\pi }\int_{-\infty }^{\infty }d\Omega \frac{%
f_{2}(z_{-},z_{+})}{\left[ 1-\nu f_{1}(z_{-})\right] \left[ 1-\nu
f_{1}(z_{+})\right] }  \label{SumResult}
\end{equation}
from which we obtain the time dependent nuclear magnetization by 
\begin{equation}
M_{x}(2\tau )={\cal L}^{-1}\left( \overline{M}_{x}(s)\right) _{t=2\tau },
\label{EchoResult}
\end{equation}
where ${\cal L}^{-1}$ is the inverse Laplace transform operator. Using Eqs. 
\ref{DefOfZ} to \ref{EchoResult} one can obtain the echo decay knowing only
the line shape or the FID function. Now let us examine three simple cases:

{\it Ising field} - this case materializes when the observed spin is coupled
only to one unobserved spin $1/2$ which fluctuates stochastically, leading
to either field up or down at the observed site. In this case the field
distribution is given by 
\[
\rho (\omega )=\frac{1}{2}\delta (\omega -\omega _{0})+\frac{1}{2}\delta
(\omega +\omega _{0}), 
\]
and its second moment is $\left\langle \omega ^{2}\right\rangle =\omega
_{0}^{2}$. Here we have to be careful about the definition of $\nu $ since
the field selection rate $\nu $, which appears in Eq.~\ref{Correlations}, is
twice the rate $\nu _{\pm }$ at which there are actual field changes. It is $%
\nu _{\pm }$ that counts in our derivation, and we find 
\begin{equation}
M_{x}^{\text{I}}(2\tau )=\frac{\exp (-2\nu _{\pm }\tau )}{f_{\text{I}}^{2}}%
\left( \omega _{0}^{2}+\nu _{\pm }f_{\text{I}}\sin (2f_{\text{I}}\tau )-\nu
_{\pm }^{2}\cos (2f_{\text{I}}\tau )\right) ,  \label{IsingResult}
\end{equation}
where $f_{\text{I}}^{2}\equiv {\omega _{0}^{2}-\nu _{\pm }^{2}}$, and the
superscript I stands for Ising. For $f^{2}<0$ the result is the same except
that $f_{\text{I}}\rightarrow i\left| f_{\text{I}}\right| $. The expansion
of Eq.~\ref{IsingResult} to lowest order in $\tau $ agrees with Eq.~\ref
{EarlyTime}. In Fig.~\ref{ThreeDist}a we present $\overline{M}_{x}^{\text{I}%
} $ on a semi-log scale as a function of $2\omega _{0}\tau $ for various
values of $\nu /\omega _{0}$. It is clear from this figure that when either $%
\nu /\omega _{0}\ll 1$ or $\nu /\omega _{0}\gg 1$ the echo decay rate is
weak compared to $\nu /\omega _{0}\simeq 1$. To quantify this phenomena we
define $T_{2}$ as the time at which the echo size decreases to $1/e$. In the
Ising case we find that $T_{2}$ reaches its minimal value of $3.146/\omega
_{0}$ when $\nu /\omega _{0}=0.69$ .

In the inset of Fig.~\ref{Param} we depict $\overline{M}_{x}^{\text{I}}$ vs. 
$(2\omega _{0}\tau )^{2}$ for $\nu /\omega _{0}=0.69$. This presentation
emphasizes a surprising fact that when $\nu /\omega _{0}\simeq 1$ the
waveform looks Gaussian over 2 orders of magnitude in echo size, even though
at early time it is exponential in $\tau ^{3}$. When fitting this Gaussian
to $M_{x}^{\text{I}}(2\tau )=\exp (-(2\omega _{0}\tau /T_{2G})^{2}/2)$ we
find $T_{2G}=2.17$.

{\it A distribution with a second moment} - It is useful to examine a
continuous distribution with a finite second moment so as to compare with
Eq.~\ref{EarlyTime}. One such distribution is: 
\begin{equation}
\rho (\omega )=\frac{2\sigma ^{3}}{\pi \left( \sigma ^{4}+4\omega
^{4}\right) }.  \label{OphirsDistr}
\end{equation}
and its second moment is given by $\left\langle \omega ^{2}\right\rangle
=\sigma ^{2}/2.$ This leads to 
\begin{eqnarray}
M_{x}(2\tau ) &=&{\frac{\,\sigma ^{2}\,e^{-2\nu \tau }}{{}\,\left( \sigma
-\nu \right) \left( \sigma -2\nu \right) }}-{\frac{\nu \sigma ^{2}e^{-\left(
\,\sigma +\nu \right) \tau }}{2(\sigma -\nu )f_{\sigma }^{2}}}  \label{OR} \\
&&-{\frac{\nu \left( \sigma ^{2}-3\nu \sigma -2{{\nu }^{2}}\right)
e^{-\left( \sigma +\nu \right) \tau }}{\,4\left( \sigma -2\nu \right)
f_{\sigma }^{2}}}\cos (2f_{\sigma }\,\tau )-{\frac{\nu \,\left( \sigma +2\nu
\right) e^{-\left( \sigma +\nu \right) \tau }}{2\,{}\,f_{\sigma }\left(
\sigma -2\,\nu \right) }}\,\sin (2f_{\sigma }\,\tau ),  \nonumber
\end{eqnarray}
where $f_{\sigma }^{2}\equiv (\sigma ^{2}-2\sigma \nu -\nu ^{2})/4$. Again,
for $f_{\sigma }^{2}<0$ the result is the same, except that $f_{\sigma
}\rightarrow i\left| f_{\sigma }\right| $. An expansion of Eq.~\ref{OR}
around $\tau =0$ agrees with Eq.~\ref{EarlyTime}, thus demonstrating the
validity of our derivation once again. In Fig.~\ref{ThreeDist}b we depict
Eq.~\ref{OR} for various values of $\nu /\sigma $. It is clear that the echo
decay is strongest for $\nu =0.88\sigma \simeq \sqrt{\left\langle \omega
^{2}\right\rangle \text{.}}$ In fact, at this value of $\nu $, $T_{2}$ is
minimal and equals $5.75/\sigma $.

{\it Lorentzian distribution} - in this case the equilibrium distribution is
taken to be 
\begin{equation}
\rho (\omega )=\frac{\lambda }{\pi (\lambda ^{2}+\omega ^{2})},
\label{LorentzField}
\end{equation}
and we find 
\begin{equation}
M_{x}^{\text{L}}(2\tau )=\frac{\lambda \exp (-2\nu \tau )-\nu \exp
(-2\lambda \tau )}{\lambda -\nu },  \label{LorentzResult}
\end{equation}
where L stands for Lorentzian. This expression has interesting properties.
An expansion of Eq.~\ref{LorentzResult} around $\tau =0$ gives 
\[
M_{x}^{\text{L}}(2\tau )=1-\frac{1}{2}\lambda \nu (2\tau )^{2}+O(\tau ^{3})
\]
which means that at early enough times the relaxation shape is Gaussian. One
should note that this expansion does not contradict Eq.~\ref{EarlyTime}
since a Lorentzian does not have a second moment. However, for $\lambda \gg
\nu $ the relaxation is exponential for $\lambda \tau \gg 1$ with the
relaxation rate $\nu $. Similarly, when $\lambda \ll \nu $ the relaxation is
exponential for $\nu \tau \gg 1$ with the relaxation rate $\lambda $. This
suggests that experimental data which stem from Eq.~\ref{LorentzResult} can
be well fitted to a stretched exponential 
\begin{equation}
M_{x}(2\tau )=\exp \left( -\left[ \frac{2\tau }{T_{2}}\right] ^{\beta
}\right)   \label{StrExp}
\end{equation}
with $1<\beta <2$. In Fig.~\ref{ThreeDist}c we depict three data sets of $%
M_{x}^{\text{L}}(2\tau )$ obtained from Eq.~\ref{LorentzResult} for various
values of $\nu /\lambda $. Unlike in the previous cases, the Lorentzian case
shows a continuous increase in relaxation rate with increasing $\nu $. In
this figure we also demonstrate the best fit of the data sets to Eq.~\ref
{StrExp}. The fits are quite good over more than an order of magnitude in
echo size, and when experimental data are fitted, Eq.~\ref{LorentzResult}
can easily be confused with Eq.~\ref{StrExp}. In Fig.~\ref{Param} we show
the parameters $\beta $, and $1/(\lambda T_{2})$ as a function of $\nu
/\lambda $. While $T_{2}$ decreases monotonically with increasing
fluctuation rate, the power $\beta $ goes through a maximum at $\nu /\lambda
=1$. However, it should be mentioned that the value of $\beta $ depends on
the range which is used for the fit.

It is interesting to compare our Lorentzian result with that of KA. In the
KA model the field dynamics at the site of the observed nuclei is generated
by flipping some other unobserved individual spins. Therefore, in their
model, it is more likely to undergo small field changes than large ones. The
situation KA tried to describe could still be approximated by the strong
collision model if $\nu \gg \lambda $ since then many unobserved spins are
flipped before the observed nuclei evolve considerably with time. This
suggests that in reality, for $\nu \gg \lambda $, we should expect $\beta =1$%
, as found here, and for $\nu \simeq \lambda $ we should expect $\beta =2$
as found by KA. Between these two limits $\beta $ should change continuously.

We thus provide a recipe for obtaining the time dependence of the echo size
for a given field distribution. We examined three particular cases and found
a natural explanation for experimental and conceptual features, such as
stretched exponential relaxation and $T_{2}$ minima, which have not been
explained quantitatively before.


\begin{figure}[tbp]
\caption[1]{A demonstration of the indirect echo when there are no local
field fluctuations (a), and when the field is dynamical (b) and changes
instantaneously.}
\label{EchoDemo}
\end{figure}

\begin{figure}[tbp]
\caption[1]{The echo decay for: (a) the Ising field (Eq.~\ref{IsingResult})
plotted against $2\protect\omega_0\protect\tau$, (b) A continuouse
distribution with a second moment (Eq.~\ref{OR}) vs. $2\protect\sigma\protect%
\tau$, and (c) Lorentzian field distribution (Eq.~\ref{LorentzResult}) as a
function of $2\protect\lambda\protect\tau$. The solid line in panel (c)
represents a fit to Eq.~\ref{StrExp} as described in the text.}
\label{ThreeDist}
\end{figure}

\begin{figure}[tbp]
\caption[1]{The parameters $1/(\protect\lambda T_2)$ and $\protect\beta$
which allow the best approximation of Eq.~\ref{LorentzResult} with Eq.~\ref
{StrExp}. In the inset, the Echo decay for the Lorentzian case with $\protect%
\nu/\protect\lambda=0.7$ is shown. }
\label{Param}
\end{figure}

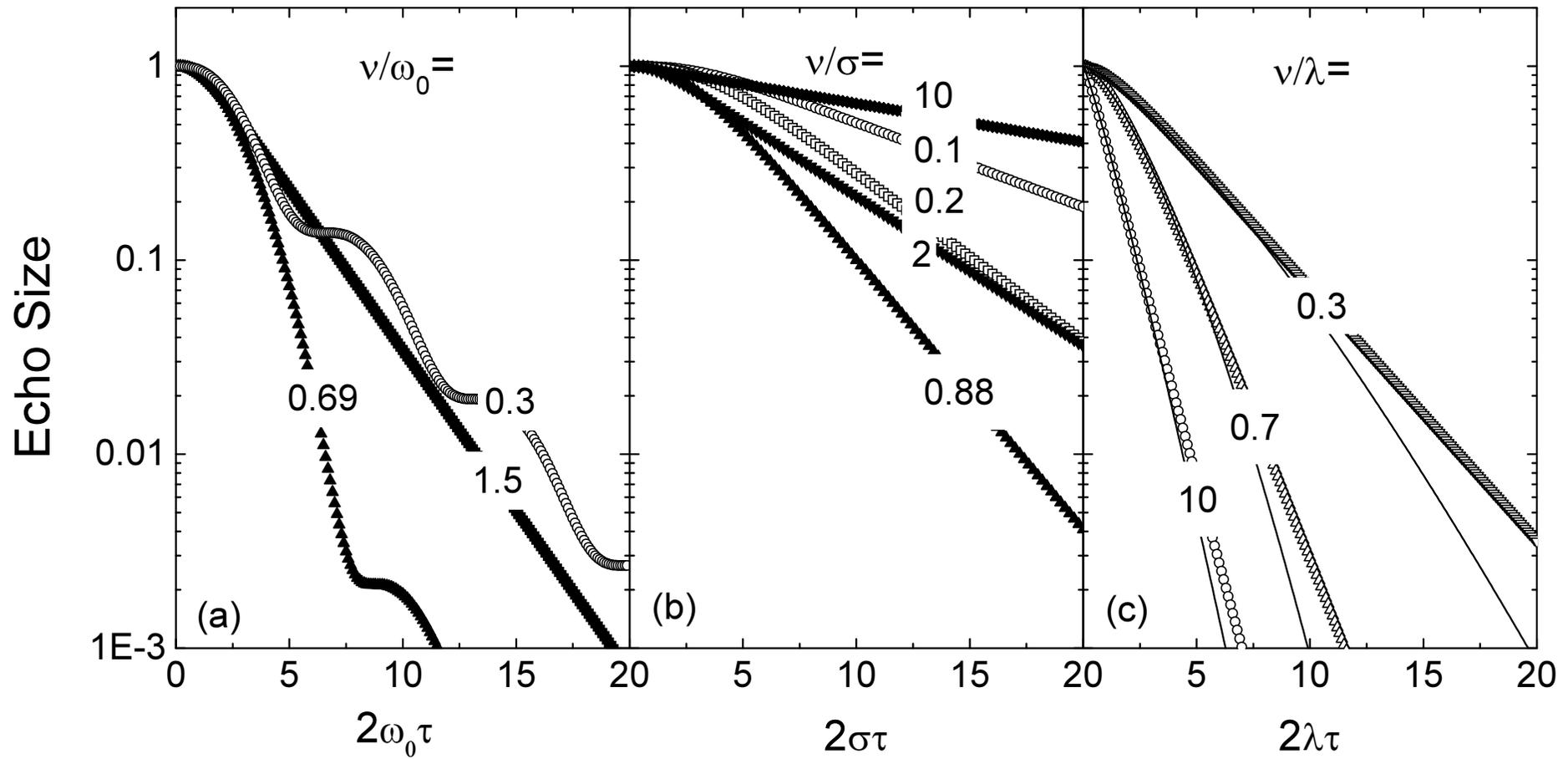
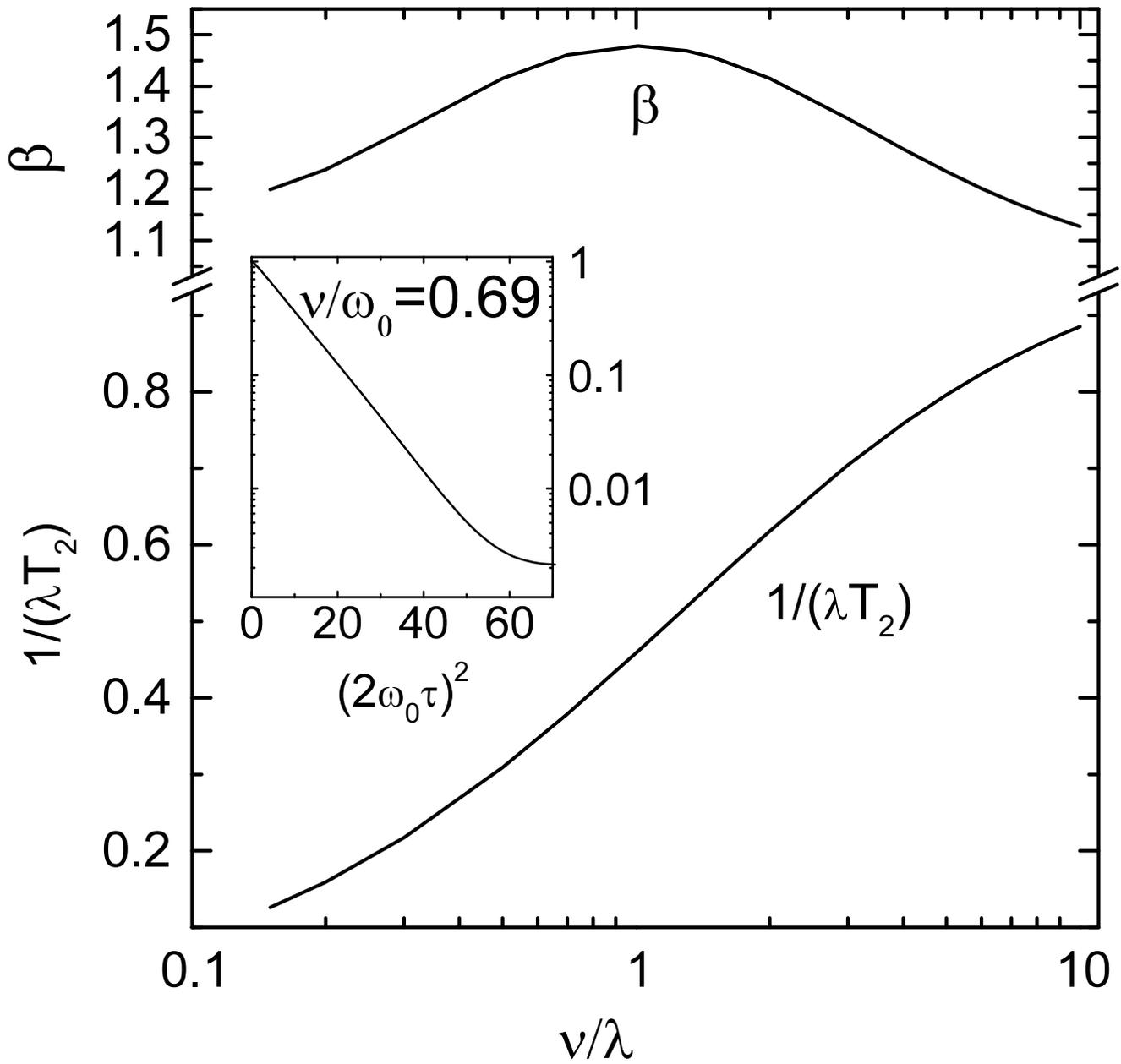
\end{document}